\documentclass[aps,amsmath,amssymb,preprintnumbers,nofootinbib]{revtex4-1}
\usepackage{amsmath,amssymb}
\usepackage{graphicx}
\usepackage{multirow}
\usepackage{subfigure}
   \usepackage{physics} 
\usepackage{acronym}
\usepackage{hyperref}
\usepackage{fancyhdr,lipsum}

\newcommand{\beq}{\begin{equation}}
\newcommand{\eeq}{\end{equation}}
\newcommand{\bea}{\begin{eqnarray}}
\newcommand{\eea}{\end{eqnarray}}
\newcommand{\be}{\begin{equation}}
\newcommand{\ee}{\end{equation}}

\begin{document}

\title{Evidence for complex fixed points in pandemic data}

\author{Giacomo Cacciapaglia$^{\ast}$} 
\affiliation{\mbox{Institut de Physique des deux Infinis de Lyon (IP2I),  UMR5822, CNRS/IN2P3, F-69622, Villeurbanne, France}}
\affiliation{\mbox{University of Lyon, Universit{\' e} Claude Bernard Lyon 1,  F-69001, Lyon, France}}

\author{Francesco Sannino}
\affiliation{CP3-Origins \& the Danish Institute for Advanced Study, University of Southern Denmark, Campusvej 55, DK-5230 Odense, Denmark}
\affiliation{ Dipartimento di Fisica E. Pancini, Universit\`a di Napoli Federico II \& INFN sezione di Napoli, Complesso Universitario di Monte S. Angelo Edificio 6, via Cintia, 80126 Napoli, Italy}

\begin{abstract}
 Epidemic data show the existence of a region of quasi-linear growth  (strolling period) of infected cases extending in between waves.   
We demonstrate that this constitutes evidence for the existence of near time-scale invariance that is neatly encoded via complex fixed points in the epidemic Renormalisation Group approach. As a result we achieve a deeper understanding of  multiple wave dynamics and its inter-wave strolling regime. Our results are tested and calibrated against the COVID-19 pandemic data. Because of the simplicity of our approach that is organised around symmetry principles our discovery amounts to a paradigm shift in the way epidemiological data are mathematically modelled. 
\end{abstract}

\maketitle

{\bf 
Pandemics are a threat to humanity and therefore understanding their spreading dynamics is paramount to controlling it. The disease diffusion dynamics is traditionally described via compartmental models \cite{Kermack:1927} or complex network diffusion techniques \cite{PERC20171,WANG20151,WANG20161},  providing an accurate description of the initial time evolution of the number of affected individuals. Another symmetry based approach is the epidemic Renormalisation Group (eRG) framework \cite{DellaMorte:2020wlc,Cacciapaglia:2020mjf,mcguigan2020pandemic}, shown to   provide robust prognoses for the time evolution of a pandemic across different regions of the world.

An extremely important period for any pandemics is the one bridging two waves, whose striking feature is a {\it strolling}  increase of infections. It is a challenge to consistently model {\it strolling} as part of the inter-wave dynamics within the current approaches.

Here we demonstrate that {\it strolling} data constitute evidence for the existence of near time-scale invariance that is efficiently encoded in complex fixed points of the eRG beta function. As a result we achieve an economic and profound understanding of the wave dynamics and the bridging period between waves. COVID-19 pandemic data are used to confirm, test and calibrate the complex eRG framework (CeRG). Because of the simplicity of the CeRG  that is organised around symmetry principles the discovery amounts to a paradigm shift in the way epidemiological data are mathematically modelled, classified and understood.  The CeRG or {\it strolling pandemic framework} is applicable to a wide range of infectious disease dynamics because it provides the bedrock of consistent mathematic modelling based on symmetry principles. It also offers a guiding principle to unveil the underlying microscopic dynamics. 
}

\vskip 1cm

Pandemics are becoming a growing threat to our society~\cite{Morense00812-20}, with  COVID-19  being the latest example \cite{Perc2020,Zhou2020,Hancean2020}. It is therefore of paramount importance to understand the diffusion of the virus in order to design effective protocols to control its spreading in the population \cite{Lai2020,Flaxman2020,Chinazzi,scala2020}.
Data collected in various instances show that the number of infected people in a limited region grows exponentially at the beginning, while then subsiding after a period of time characteristic of each type of virus. This feature can be effectively described by various mathematical models, including compartmental ones \cite{Kermack:1927}, complex network diffusion techniques \cite{PERC20171,WANG20151,WANG20161} and the eRG approach \cite{DellaMorte:2020wlc,Cacciapaglia:2020mjf,mcguigan2020pandemic}. Going beyond the first wave is however a challenge \cite{Scudellari}.

Epidemic data for multi-wave pandemics show the existence of a region of quasi-linear growth of infected cases extending in between different waves. In this period of time, the number of new infected cases grows much slower than the exponential  growth in each wave, thus  
we refer to it as the region of {\it strolling} epidemic regime. 
    The scope of our work is to demonstrate that:
    \begin{itemize} 
    \item[i)] understanding the strolling regime is important to achieve a deep understanding of the underlying epidemic dynamics, in a unified way within and between waves; 
    \item[ii)] near time-scale invariance is key to such an understanding; 
    \item[iii)] the eRG approach \cite{DellaMorte:2020wlc,Cacciapaglia:2020mjf}, when extended to include complex fixed points, is the ideal framework to explain and model strolling dynamics. 
    \end{itemize} 
  Evidence for the above comes from  applying the novel framework to COVID-19 data in Europe and in the US. Here, strolling dynamics eminently explains the pandemic diffusion data showing that the novel approach achieves a better characterisation of the data compared to the original eRG with real fixed points. Thus the resulting framework constitutes a paradigm shift in our understanding of epidemic diffusion dynamics.
 
    The  eRG framework  \cite{DellaMorte:2020wlc,Cacciapaglia:2020mjf,mcguigan2020pandemic} is based on a single differential equation describing the time evolution of the total number of infected cases, inspired by particle physics methods \cite{Wilson:1971bg,Wilson:1971dh}. It has been shown to be highly effective when describing how the pandemic spreads across different regions of the world \cite{Cacciapaglia:2020mjf}, and to be able to effectively predict the time frame of  a second wave~\cite{cacciapaglia2020second}.  Due to the presence of real fixed points, in the original eRG approach the onset of the second wave had to be modelled independently alongside the region bridging the waves. The link between the eRG and traditional compartmental approaches  \cite{Kermack:1927} can be found in \cite{DellaMorte:2020qry}.  In this article we propose the novel {\it strolling} paradigm as a unified way to model and understand epidemic data, within and between waves, that stems from the emergence of complex zeros of the eRG beta function.  
        
        The strolling regime in epidemiology has an important counterpart in particle and condensed matter physics. It has to do with the loss of near-scale invariance. In particle physics, the latter is married to special relativity, thus yielding what is known as conformal invariance. 
        Depending on the underlying mechanism behind the loss of conformality, one can envision several scenarios ranging from a Berezinski--Kosterlitz--Thouless (BKT)-like phase transition, first discovered in two dimensions~\cite{Kosterlitz:1974sm} and later proposed for four dimensions in~\cite{Miransky:1984ef,Miransky:1996pd,Holdom:1988gs,Holdom:1988gr,
Cohen:1988sq,Appelquist:1996dq,Gies:2005as},  to a jumping (non-continuous) phase transition~\cite{Sannino:2012wy}. 
Evidence for the BKT transition has been found in various two-dimensional materials and physical systems \cite{BKT1,BKT2,BKT3,BKT4}.
A large body of numeric and analytic work has followed the discovery that  one  can achieve (near) conformal dynamics in four dimensions with small number of matter fields \cite{Sannino:2004qp}. The work culminated in the well-known conformal window phase diagram~\cite{Dietrich:2006cm} that has served as a road map for first-principle lattice  studies, as summarised in~\cite{Cacciapaglia:2020kgq}.           
        
        \vskip .2cm
In this work we demonstrate that an approximate scale-invariant dynamics  can explain the emergence of the strolling regime as well as of a new wave, in epidemiological data.  Our approach is organised around symmetry principles, which allow for a compact and efficient way to analyse the data. Our discovery amounts to a paradigm shift in the way epidemiological data will be modelled in the future. This discovery offers also a precious guideline in unveiling underlying models aimed at a microscopic understanding of multi-wave diffusion of the pandemics, inspired by the work done in particle and condensed matter physics.
   
   \vskip .2cm
The key equation of the {\it strolling} framework is the complex eRG (CeRG, pronounced as {\it Serge}) beta function (we will explain it in more detail in the section Methods):
   \begin{equation}
- \beta_{\rm CeRG} (\alpha) = \frac{d \alpha}{dt} =   \alpha  \left[  \left(1 - {\alpha} \right)^2  - \delta\right]^p \ .
 \label{eq:beta2intro}
\end{equation}
The solutions of this differential equation, $\alpha (t)$, will be used to characterise the time-evolution of the number of infected cases.
This beta function features the following zeros: 
\begin{equation}
 \alpha_0 =0 \ , \qquad \alpha_{1} =  1 - \sqrt{\delta} \ , \qquad \alpha_{2} =  1 +  \sqrt{\delta} \ ,
 \end{equation} 
corresponding to time-scale invariant fixed points of the theory. This implies that if $\alpha$ equals any of these values, at any given time, its value will remain {\it fixed}.  For positive $\delta$ and an initial value of $\alpha$ in between zero and $\alpha_1$, the solution interpolates between zero and the first fixed point. This dynamics is the one employed in the first eRG studies \cite{DellaMorte:2020wlc,Cacciapaglia:2020mjf,mcguigan2020pandemic}, where $p$ was  $1/2$ with $\delta = 0$. It nicely encodes the time-scale invariance at short and large times, as well as the fast exponential growth in between the first two zeros.  

\begin{figure}
         \includegraphics[width=16cm]{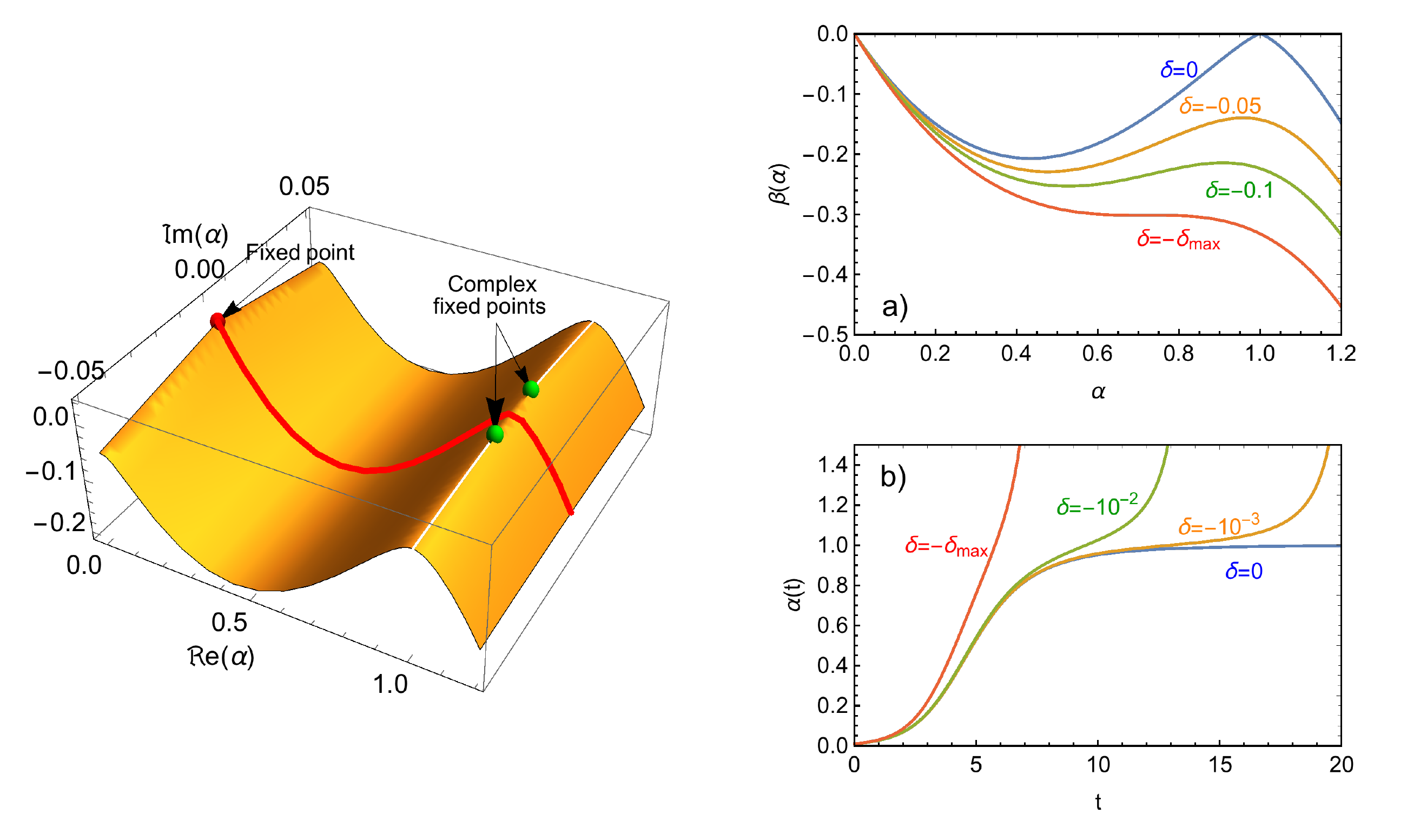} 
        \caption{Emergence of the \emph{strolling} dynamics from the CeRG beta function. Left panel: illustration of the beta function ($-|\beta (\alpha)|$) extended to the complex plane, i.e. considering a complex $\alpha$. The red line represents the trajectory on the real plane, emerging from the real fixed point at $\alpha = 0$ (red dot): the strolling emerges as the solution slows down when passing between the two complex fixed points (green dots). Right panel: a) beta functions and b) solutions for $p=0.65$ and various choices for $\delta$.}
   \label{fig:1}
\end{figure}

For negative $\delta$ the two non-trivial fixed points become complex and therefore can't be reached. Nevertheless the dynamics still feels their presence for sufficiently small $|\delta|$. The overall effect is that the solution spends much time near the would-be fixed point at $\alpha = 1$, where it  features a slow linear rise.  This behaviour is naturally identified with the {\it strolling} regime. This dynamics is illustrated in the left plot of Fig.~\ref{fig:1}, where we display the beta function analytically continued in the complex plane as function of a complex $\alpha$ (more precisely, we plot minus its absolute value). The trajectory of $\alpha$, on the real plane, is indicated by the red line, which originates at the real fixed point $\alpha_0 = 0$.  The valley with negative beta function values drives the exponential growth of $\alpha$ with time, until the near fixed point value is reached, at the local maximum $\alpha \approx 1$. The trajectory, thus, needs to pass through the two fixed points in the complex plane, indicated by the green dots: the closer they are to the real plane (i.e., the smaller $|\delta|$), the slower will be the evolution with time and the longer time $\alpha$ will spend close to the complex fixed points. This is shown in the panel b), where we plot the solutions for $p=0.65$ and various values of $\delta$. The value $\delta_{\rm max} \simeq 0.18 $ corresponds to the largest value of $|\delta|$ after which the real-axes valley disappears, as shown in panel a).  A more in depth analysis on the properties of the CeRG beta function and of the solutions are reported in the section Methods.

The solutions of \eqref{eq:beta2intro} constitute a two parameter family of functions that we use to efficiently model the COVID-19 epidemic data. To do so we identify 
   $\alpha(t)$ with the logarithm of the number of infected cases up to a normalisation and $t$ is the time variable rescaled by a constant $\gamma$ measured in weeks.

\section*{Results}

\begin{figure}
         \includegraphics[width=16cm]{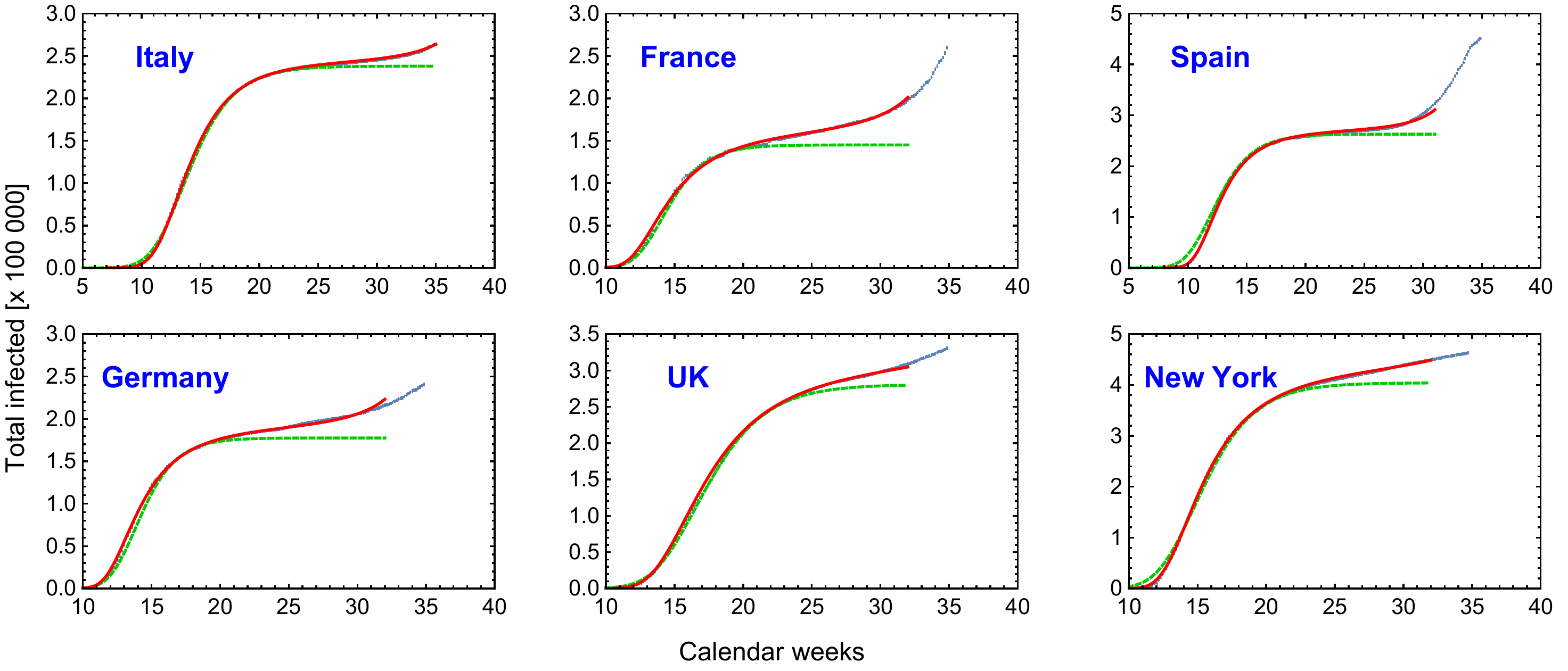} 
        \caption{Fit of the CeRG solutions (red curves) compared to the total number of infected cases (blue dots) adjourned to the 28th of August. For comparison, in dashed green we show the eRG first wave fits obtained in \cite{cacciapaglia2020second}, which do not feature the strolling dynamics. The epidemiological data is from \href{www.worldometer.info}{www.worldometer.info}.}
   \label{fig:fits}
\end{figure}

The solutions of Eq.~\eqref{eq:beta2intro} contain five parameters that can be fitted on the epidemiological data: the two parameters characterising the family of solutions, $p$ and $\delta$, two normalisation factors $a$ and $\gamma$, and the initial condition. The latter determine the beginning of the infection spread. The normalisations appear as
\begin{equation}
\alpha = \frac{1}{a} \ln \mathcal{I}\,, \qquad t = \gamma t_w\,,
\end{equation}
where $t_w$ is the time measured in weeks, and $\mathcal{I}$ is the total number of infected in each region that we consider. Note that the infection rate $\gamma$ and the normalisation $a$ are equivalent to the parameters of the original eRG approach, for $p=1/2$ and $\delta=0$.
The results for 6 countries/regions is shown in Fig.~\ref{fig:fits}, where the blue dots indicate the data (from \href{www.worldometer.info}{www.worldometer.info}) while the red curve is the solution of the CeRG model. We choose to test the model against Italy, France, Spain, Germany, the UK and New York state for two reasons: these regions already feature the end of the first wave and the {\it strolling} regime before the beginning of the second wave; the number of cases is large enough to provide a good statistics with a consistent testing practice. The data show that the presence of the {\it strolling} regime is a physical property of the pandemic.

The figure clearly shows that the CeRG model provides an excellent description of the data. As a comparison, in dashed green we also show the fit from the original eRG model, which has only 3 parameters. The values of the parameters used in the plots are listed in Table~\ref{T1}. The effect of each parameter can be easily understood: $a$ determines the overall normalisation of the number of cases (more precisely of its natural logarithm), while the infection rate $\gamma$ makes the exponential growth in the curve more or less steep by rescaling the time. The new parameter $p$ smoothens the curve when it approaches the near-fixed point at $\alpha \approx 1$. Thus, tuning $p$ allows to improve the fit of the exponentially growing initial phase, i.e. the first wave. The role of $\delta$ is to determine the flatness of the {\it strolling} phase, namely the constant number of new cases registered after the end of the first wave, and the time when the second exponential growth begins. It is, therefore, non-trivial to be able to reproduce both with a single solution. Once the second exponential phase starts, the model looses validity, because the solution diverges (we stop the red curved at this stage).

\begin{table}
\begin{tabular}{ || l | c|c|c|c||}
\hline
 & \multicolumn{4}{c||}{eRG and CeRG parameters}\\
\hline
 & $\gamma$ & $a$ & $p$ & $\delta$ \\
\hline
Italy (CeRG) & $0.69$ & $12.40$ & $0.58$ & $3.1 \cdot 10^{-6}$ \\
Italy (eRG) & $0.43$ & $12.38$ & $0.5$ & -- \\ \hline
France (CeRG) & $1.0$ & $11.98$ & $0.647$ & $4.0\cdot 10^{-5}$ \\
France (eRG) & $0.584$ & $11.89$ & $0.5$ & -- \\ \hline
Spain (CeRG) & $1.0$ & $12.51$ & $0.60$ & $4.0\cdot 10^{-6}$ \\
Spain (eRG) & $0.53$ & $12.48$ & $0.5$ &  -- \\ \hline
Germany (CeRG) & $1.1$ & $12.15$ & $0.635$ & $1.4\cdot 10^{-5}$ \\
Germany (eRG) & $0.616$ & $12.09$ & $0.5$ & -- \\ \hline
UK (CeRG) & $0.70$ & $12.63$ & $0.64$ & $3.0\cdot 10^{-5}$ \\
UK (eRG) & $0.368$ & $12.55$ & $0.5$ & -- \\ \hline
New York (CeRG) & $0.95$ & $12.98$ & $0.65$ & $1.9\cdot 10^{-5}$ \\
New York (eRG) & $0.42$ & $12.91$ & $0.5$ &  -- \\ 
\hline
 \end{tabular}
  \caption{CeRG and eRG parameters used to obtain the curves in Fig.~\ref{fig:fits}.}
 \label{T1}
 \end{table}

\begin{figure}
         \includegraphics[width=8cm]{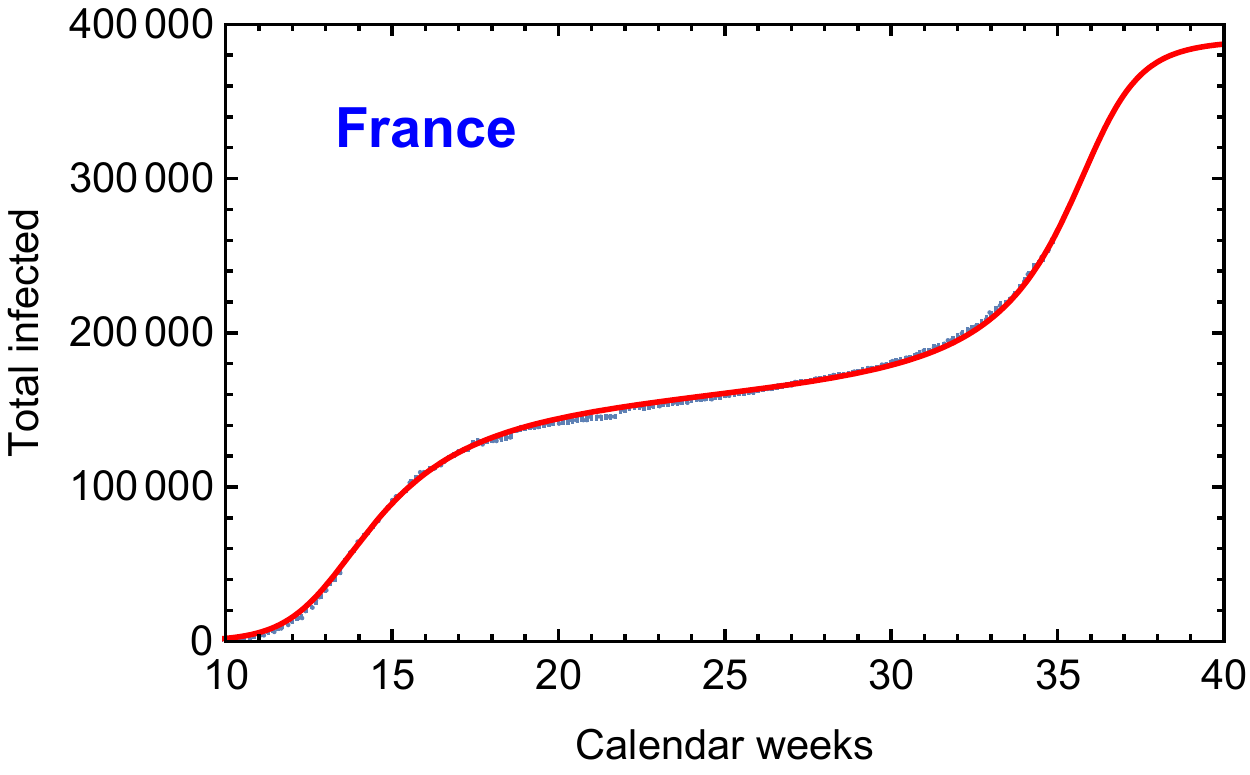} 
        \caption{Fit of the second wave model for France. The parameters we use to draw the red curve are: $\gamma = 1.15$, $a = 1.6 \cdot 10^5 = e^{11.98}$, $\zeta = 0.41$, $p=0.75$, $p_2 = 1.2$. The epidemiological data is from \href{www.worldometer.info}{www.worldometer.info}.}
   \label{fig:2nd}
\end{figure}

Having reproduced the {\it strolling} phase and the time of the second wave onset, it is tantalising to extend the model to be able to fully fit the second wave.
The easiest way would be to endow the CeRG beta function in Eq.~\eqref{eq:beta2intro} with another zero as follows:
   \begin{equation}
- \beta_{\rm 2-waves} (\alpha) = \frac{d \alpha}{dt} =   \alpha  \left[  \left(1 - {\alpha} \right)^2  - \delta\right]^p \; \left( 1-\zeta \alpha \right)^{p_2}\ ,
 \label{eq:beta2wave}
\end{equation}
with $\zeta < 1$. The additional factor introduces a second real fixed point at $\alpha_4 = 1/\zeta > 1$, so that the solution will flow to it after the second exponential growth starts. However, because we are using the log of the infected cases, the would-be first fixed point and the second one occur at values too close to each other, i.e. $\zeta \approx 1$. Therefore the additional factor in \eqref{eq:beta2wave} makes it difficult to reproduce the first wave and the strolling regime. A possible way out is to directly identify the function $\alpha$ with the number of infected cases rather than its logarithm, i.e. $\alpha = \mathcal{I}/a$. In this case we are able to fit the data as  Fig.~\ref{fig:2nd} 
for France clearly shows. The resulting prognosis is that France will double the total number of infected individuals by week 40, i.e. by the end of September. At this point we find ourself in the following conundrum: without the full description of the second wave the CeRG beta function for the logarithm of the infected cases is a better description of the initial wave and subsequent strolling phase than the same-type beta function written directly for $I(t)$, however a two-wave analysis is well described by the CeRG beta function that includes another zero and interpreted directly for $I(t)$.  

\section*{Discussion}

We provided a new physical paradigm for describing pandemic dynamics, which is able to naturally account for the observed strolling phase in between waves. It is based on the realisation that the strolling phase appears as a manifestation of near time-scale invariance of the underlying pandemic diffusion theory. In the CeRG framework this is encoded in the emergence of complex fixed points of its beta function.  
The discovery is supported by the COVID-19 data that we accurately reproduce with solutions of the CeRG beta function. 

In particular the approach correctly describes, in a better way than the eRG, the exponential growth of the first wave. This is so because it allows for larger values of $\gamma$, as shown in Table ~\ref{T1}, which better reproduce the initial data of the exponential epidemic growth. The fact that the initial data needed a larger  infection rate $\gamma$ was already observed in early data fits \cite{DellaMorte:2020wlc}. Additionally, allowing the parameter $p$ to be greater than $1/2$ slows the epidemic curve near the ending of the first wave, again in better agreement with data. Strolling was not part of the previous approach and here it depends on the parameter $\delta$, which carries the physical significance of controlling the distance from exact time-dilation invariance. The numerical value of $\delta$ determines both the slope of the strolling (the constant number of new infected cases after the first wave) and its duration before the onset of the second wave. Thus the CeRG solution predicts the beginning of the second wave once we know the strolling slope. Caveats apply to this prediction power because of the presence of additional effects that a single equation cannot embody, such as interactions across different regions of the world. These interactions have been shown to be important for the beginning of the second wave \cite{Cacciapaglia:2020mjf}. This offers a natural solution to the conundrum discussed at the end of the previous section, i.e. to use the CeRG approach to describe one wave and its subsequent strolling regime and repeat it for any subsequent wave including the interactions with other regions of the world as done in \cite{Cacciapaglia:2020mjf}.  

The CeRG approach, which is based on the implementation of important symmetries of the pandemic, is a macroscopic realisation that serves as a guiding principle to unveil its microscopic dynamics. A natural step in this direction is to consider a possible realisation in terms of a BKT-like theory \cite{Kosterlitz:1974sm}. 
 
Our discovery amounts to a paradigm shift in the way epidemiological data are mathematically modelled, classified and understood, and we further believe that the CeRG  framework can be applied to a wide range of diffusion-based social dynamics.

\section*{Methods}
 
\subsection{Review of the epidemic Renormalisation Group}
In the original \emph{eRG} approach \cite{DellaMorte:2020wlc}, rather than the number of cases, it was used its natural logarithm  
$\alpha(t) = \ln {I}(t)$. For a single wave pandemic this provides a better fit to  the data than for a similar equation for $I(t)$ \cite{mcguigan2020pandemic}. 
The derivative of $\alpha(t)$ with respect to time provides a new quantity that we interpret as the {\it beta function} of an underlying microscopic model. In statistical and high energy physics, the latter governs the time (inverse energy) dependence of  the interaction strength among fundamental particles. Here it regulates infectious interactions.

 More specifically, as the renormalisation group equations in high energy physics are expressed in terms of derivatives with respect to the energy $\mu$, it is natural to identify the time as
  $t/t_0=-\ln ({\mu/\mu_0})$, where $t_0$ and $\mu_0$ are respectively a reference time and energy scale. We choose $t_0$ to be one week so that time is measured in weeks, and will drop it in the following. 
 Thus, the dictionary between the eRG equation for the epidemic strength $\alpha(t)$ and the high-energy physics analog is
 \begin{equation}
 \label{betaalpha}
 \beta (\alpha(t)) = \frac{d \alpha(t)}{d\ln\left(\mu/\mu_0\right)} = - \frac{d \alpha(t)}{dt} \ . 
\end{equation}  
The pandemic beta function able to represent an isolated region of the world  \cite{DellaMorte:2020wlc} can be parametrised as
\beq
- \beta (\alpha) = \frac{d \alpha(t)}{dt} = {\gamma} \, \alpha   \left( 1 - \frac{\alpha}{a} \right)^{2p}\,,
\label{eq:beta0}
\eeq 
 whose solution, for $2p=1$, is a familiar logistic-like function
\begin{equation}
\alpha (t) =  \frac{a e^{{\gamma} t}}{b + e^{{\gamma} t}}\,.
\end{equation}
The dynamics encoded in Eq.~\eqref{eq:beta0} is that of a system that flows from an Ultra-Violet fixed point at $t=-\infty$ where $\alpha = 0$  to an Infra-Red one where $\alpha = a$. The latter value encodes the total number of infected cases  in the region under study. The coefficient ${\gamma}$ is the diffusion slope, while $b$ shifts the entire epidemic curve by a given amount of time. Further details, including what parameter influences the {\it flattening of the curve} and location of the inflection point and its properties can be found in \cite{mcguigan2020pandemic} and in \cite{DellaMorte:2020wlc}. 

The rate with which the fixed points are approached is determined by a universal quantity termed scaling exponent:
\begin{equation}
\vartheta = \frac{\partial \beta}{ \partial  \alpha} \Big |_{\alpha^\ast} \ , \quad {\rm with} \quad \alpha^\ast = {\rm fixed~point} \ .  
\end{equation}
At $\alpha^\ast = 0 $ and $\alpha^\ast = a$ we have respectively
\begin{equation}
\vartheta(0) = - \gamma \ , \qquad \vartheta(a) = \gamma \ .
\end{equation}
A negative (positive) exponent means that the fixed point is repulsive (attractive).

The presence, however, of a truly interacting fixed point at large times  predicts that the number of new cases drops to zero. This is, however, not what is observed for COVID-19 for most of the countries. The system does not reach a time-scale invariant theory.  What it is generally observed is the occurrence of a temporal region of roughly constant number of new infected cases. After this time the system, if there is no heard immunity, will start a new epidemic wave. The extent to which the system remains in this state in between two waves depends on the intrinsic dynamics of the virus as well as social distancing measures. The point we will now address is how this important phenomenon can be encapsulated in a mathematically consistent way as a controllable deformation of a symmetry limit of the model, i.e. a phenomenon emerging as near time-dilation. 

\subsection{Complex epidemic Renormalisation Group (CeRG): Strolling region of pandemics }
Here we propose the \emph{CeRG} model for which the beta function in \eqref{eq:beta0} becomes: 
\begin{equation}
- \beta (\alpha) = \frac{d \alpha}{dt} = {\gamma}\, \alpha  \left[  \left(1 - \frac{\alpha}{{a}} \right)^2  - \delta\right]^p
 = {{\gamma}} \, \alpha \left(\frac{\alpha}{{a}} - 1 +  \sqrt{\delta} \right)^p   \left(\frac{\alpha}{{a}} - 1 -  \sqrt{\delta} \right)^p \, , 
\label{eq:beta1}
\end{equation}
with $\delta$ and $p$ real numbers and $p$ positive. One can rescale the time by $1/ \gamma $ and  $\alpha$ by  $a$ per each country to eliminate  them from the equations so that we can write  the beta function in the form of Eq.~\eqref{eq:beta2intro}.
 Here we are interested in negative values of $\delta = - |\delta |$  leading to the following zeros of the beta function: 
\begin{equation}
 \alpha_0 =0 \ , \quad \alpha_{1} = 1 - i \sqrt{|\delta|} \ , \quad \alpha_{2} =  1 + i \sqrt{|\delta |} \ , 
 \end{equation} 
with the complex zeros each of order $p$. The zero at the origin corresponds to a repulsive one, meaning that it drives the beginning of the infection, while the other two zeros control the dynamics at large times, as discussed in the main text. For $\delta = 0 $ we recover the original eRG of Eq.~\eqref{eq:beta0}, featuring physical fixed points.  Since, at each of the two complex fixed points the beta function vanishes, one observes the occurrence of two complex time-dilated invariant theories.  

As shown in Fig.~\ref{fig:1}, the beta function has a local maximum at values of $\alpha \approx 1$, which becomes flatter for larger $|\delta|$, until it disappears for $\delta \leq - \delta_{\rm max} = - p^2/(1+2p)$. Thus for large negative values of $\delta$, the strolling regime is lost and the solution will keep growing exponentially fast.

For small $|\delta|$, the solutions feature a period of slow linear growth, as shown in panel b) of Fig.~\ref{fig:1}: this period we identify with the {\it strolling} regime. In this case, the solution for $\alpha \lesssim 1$ can be approximated by the beta function with $\delta = 0$, which allows for an analytic solution in terms of Hypergeometric functions 
\begin{equation}
\int_{t_i}^{t}dt = \int_{\alpha_{in}}^{\alpha \leq 1} \frac{d\alpha}{ \alpha  \left[  \left( 1 - {\alpha} \right)^2  - \delta \right]^p} 
=2 p \ \left( \alpha \, {}_3 F_2\left[ 1 ,1,1+2p;2,2; \alpha \right] - \{\alpha \rightarrow \alpha_{in}\}  \right) + \ln \frac{\alpha}{\alpha_{in}} \ .
  \end{equation}

\begin{figure}
           \includegraphics[width=8cm]{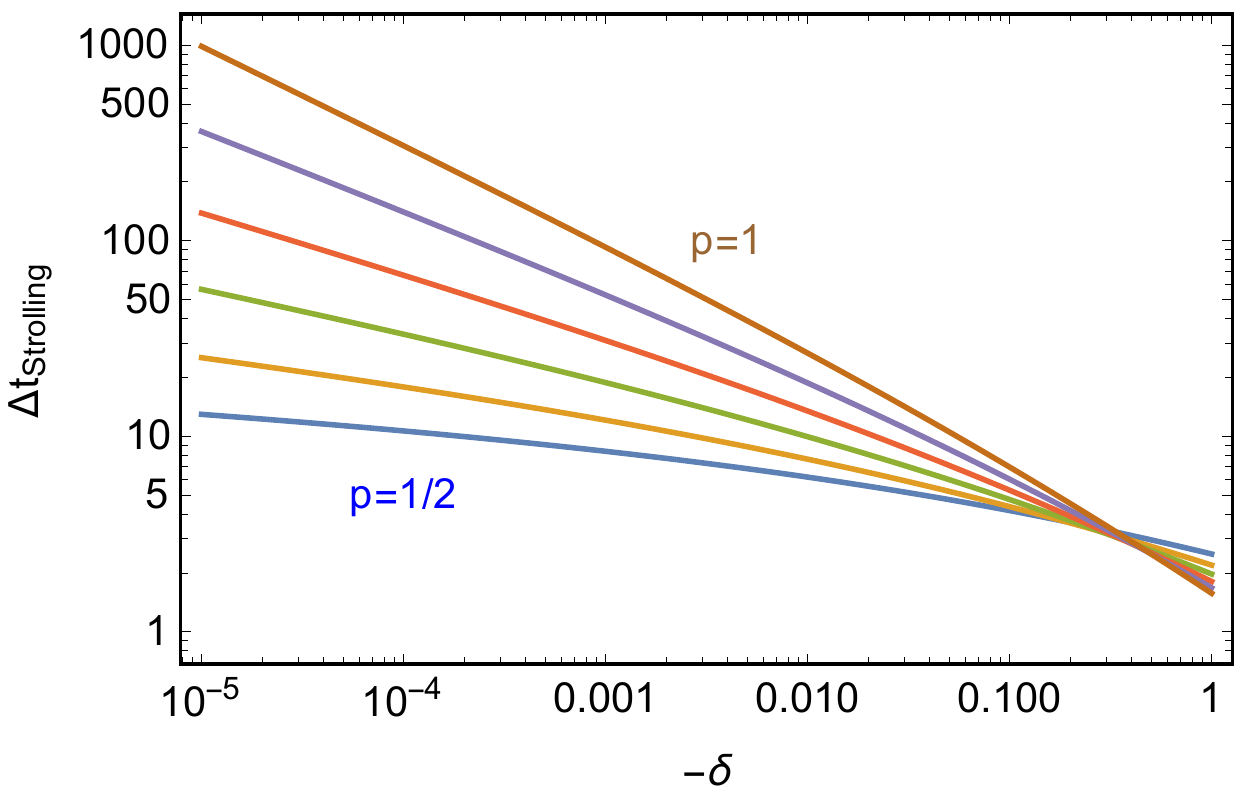} 
        \caption{Values of $\Delta t_{\rm  Strolling}$ for $p=0.5, 0.6, 0.7, 0.8, 0.9$ and $1$.}  
   \label{fig:strolling}
\end{figure}

The {\it strolling} regime occurs because the real beta function develops a maximum near $\alpha = 1$ for small $|\delta|$, technically allowing the theory to feel the nearby presence of the complex fixed points.  For any fixed $p$, the duration of the {\it strolling} region increases with decreasing $|\delta|$. 
The duration can be estimated as follows
\begin{equation}
\Delta t_{\rm  Strolling} = - 2 \int_1^\infty  \frac{d \alpha}{\beta_{\rm CeRG} (\alpha)}\ .
\end{equation}
The results as a function of $\delta$, for different values of $p$, are shown in Fig.~\ref{fig:strolling}. For fixed $p$, the strolling time grows like an inverse power law of $\delta$, while for fixed and sufficiently small $\delta$ it increases exponentially with $p$.

\section*{Data availability}
The data for the COVID-19 infected cases in Europe are extracted from the \href{www.worldometer.info}{www.worldometer.info} repository.

\bibliographystyle{naturemag}
\bibliography{biblio}

\begin{thebibliography}{10}
\expandafter\ifx\csname url\endcsname\relax
  \def\url#1{\texttt{#1}}\fi
\expandafter\ifx\csname urlprefix\endcsname\relax\def\urlprefix{URL }\fi
\providecommand{\bibinfo}[2]{#2}
\providecommand{\eprint}[2][]{\url{#2}}

\bibitem{Kermack:1927}
\bibinfo{author}{Kermack, W.~O.}, \bibinfo{author}{McKendrick, A.} \&
  \bibinfo{author}{Walker, G.~T.}
\newblock \bibinfo{title}{{A contribution to the mathematical theory of
  epidemics}}.
\newblock \emph{\bibinfo{journal}{Proceedings of the Royal Society A}}
  \textbf{\bibinfo{volume}{115}}, \bibinfo{pages}{700--721}
  (\bibinfo{year}{1927}).
\newblock
  \urlprefix\url{https://royalsocietypublishing.org/doi/10.1098/rspa.1927.0118}.

\bibitem{PERC20171}
\bibinfo{author}{Perc, M.} \emph{et~al.}
\newblock \bibinfo{title}{Statistical physics of human cooperation}.
\newblock \emph{\bibinfo{journal}{Physics Reports}}
  \textbf{\bibinfo{volume}{687}}, \bibinfo{pages}{1 -- 51}
  (\bibinfo{year}{2017}).
\newblock
  \urlprefix\url{http://www.sciencedirect.com/science/article/pii/S0370157317301424}.

\bibitem{WANG20151}
\bibinfo{author}{Wang, Z.}, \bibinfo{author}{Andrews, M.~A.},
  \bibinfo{author}{Wu, Z.-X.}, \bibinfo{author}{Wang, L.} \&
  \bibinfo{author}{Bauch, C.~T.}
\newblock \bibinfo{title}{Coupled disease--behavior dynamics on complex
  networks: A review}.
\newblock \emph{\bibinfo{journal}{Physics of Life Reviews}}
  \textbf{\bibinfo{volume}{15}}, \bibinfo{pages}{1 -- 29}
  (\bibinfo{year}{2015}).
\newblock
  \urlprefix\url{http://www.sciencedirect.com/science/article/pii/S1571064515001372}.

\bibitem{WANG20161}
\bibinfo{author}{Wang, Z.} \emph{et~al.}
\newblock \bibinfo{title}{Statistical physics of vaccination}.
\newblock \emph{\bibinfo{journal}{Physics Reports}}
  \textbf{\bibinfo{volume}{664}}, \bibinfo{pages}{1 -- 113}
  (\bibinfo{year}{2016}).
\newblock
  \urlprefix\url{http://www.sciencedirect.com/science/article/pii/S0370157316303349}.

\bibitem{DellaMorte:2020wlc}
\bibinfo{author}{Della~Morte, M.}, \bibinfo{author}{Orlando, D.} \&
  \bibinfo{author}{Sannino, F.}
\newblock \bibinfo{title}{{Renormalization Group Approach to Pandemics: The
  COVID-19 Case}}.
\newblock \emph{\bibinfo{journal}{Front. in Phys.}}
  \textbf{\bibinfo{volume}{8}}, \bibinfo{pages}{144} (\bibinfo{year}{2020}).

\bibitem{Cacciapaglia:2020mjf}
\bibinfo{author}{Cacciapaglia, G.} \& \bibinfo{author}{Sannino, F.}
\newblock \bibinfo{title}{{Interplay of social distancing and border
  restrictions for pandemics (COVID-19) via the epidemic Renormalisation Group
  framework}}.
\newblock \emph{\bibinfo{journal}{in press on Sci. Rep.}}
  (\bibinfo{year}{2020}).
\newblock \eprint{2005.04956}.

\bibitem{mcguigan2020pandemic}
\bibinfo{author}{McGuigan, M.}
\newblock \bibinfo{title}{Pandemic modeling and the renormalization group
  equations: Effect of contact matrices, fixed points and nonspecific vaccine
  waning} (\bibinfo{year}{2020}).
\newblock \eprint{2008.02149}.

\bibitem{Morense00812-20}
\bibinfo{author}{Morens, D.~M.}, \bibinfo{author}{Daszak, P.},
  \bibinfo{author}{Markel, H.} \& \bibinfo{author}{Taubenberger, J.~K.}
\newblock \bibinfo{title}{Pandemic covid-19 joins history{\textquoteright}s
  pandemic legion}.
\newblock \emph{\bibinfo{journal}{mBio}} \textbf{\bibinfo{volume}{11}}
  (\bibinfo{year}{2020}).
\newblock \urlprefix\url{https://mbio.asm.org/content/11/3/e00812-20}.
\newblock \eprint{https://mbio.asm.org/content/11/3/e00812-20.full.pdf}.

\bibitem{Perc2020}
\bibinfo{author}{Perc, M.}, \bibinfo{author}{Gori\v{s}ek~Miksi\'{c}, N.},
  \bibinfo{author}{Slavinec, M.} \& \bibinfo{author}{Sto\v{z}er, A.}
\newblock \bibinfo{title}{Forecasting covid-19}.
\newblock \emph{\bibinfo{journal}{Frontiers in Physics}}
  \textbf{\bibinfo{volume}{8}}, \bibinfo{pages}{127} (\bibinfo{year}{2020}).
\newblock
  \urlprefix\url{https://www.frontiersin.org/article/10.3389/fphy.2020.00127}.

\bibitem{Zhou2020}
\bibinfo{author}{Zhou, T.} \emph{et~al.}
\newblock \bibinfo{title}{{Preliminary prediction of the basic reproduction
  number of the Wuhan novel coronavirus 2019-nCoV}}.
\newblock \emph{\bibinfo{journal}{J Evid. Based Med.}}
  \textbf{\bibinfo{volume}{13}}, \bibinfo{pages}{3--7} (\bibinfo{year}{2020}).
\newblock \urlprefix\url{https://pubmed.ncbi.nlm.nih.gov/32048815/}.

\bibitem{Hancean2020}
\bibinfo{author}{H\^{a}ncean, M.-G.}, \bibinfo{author}{Perc, M.} \&
  \bibinfo{author}{Juergen, L.}
\newblock \bibinfo{title}{Early spread of covid-19 in romania: imported cases
  from italy and human-to-human transmission networks}.
\newblock \emph{\bibinfo{journal}{R. Soc. open sci.}}
  \textbf{\bibinfo{volume}{7}}, \bibinfo{pages}{200780} (\bibinfo{year}{2020}).

\bibitem{Lai2020}
\bibinfo{author}{Lai, S.} \emph{et~al.}
\newblock \bibinfo{title}{Effect of non-pharmaceutical interventions for
  containing the covid-19 outbreak in china}.
\newblock \emph{\bibinfo{journal}{Nature}}  (\bibinfo{year}{2020}).

\bibitem{Flaxman2020}
\bibinfo{author}{Flaxman, S.} \emph{et~al.}
\newblock \bibinfo{title}{Estimating the effects of non-pharmaceutical
  interventions on covid-19 in europe}.
\newblock \emph{\bibinfo{journal}{Nature}}  (\bibinfo{year}{2020}).

\bibitem{Chinazzi}
\bibinfo{author}{Chinazzi, M.} \emph{et~al.}
\newblock \bibinfo{title}{The effect of travel restrictions on the spread of
  the 2019 novel coronavirus (covid-19) outbreak}.
\newblock \emph{\bibinfo{journal}{Science}} \textbf{\bibinfo{volume}{368}},
  \bibinfo{pages}{395--400} (\bibinfo{year}{2020}).
\newblock \urlprefix\url{https://science.sciencemag.org/content/368/6489/395}.
\newblock
  \eprint{https://science.sciencemag.org/content/368/6489/395.full.pdf}.

\bibitem{scala2020}
\bibinfo{author}{Scala, A.} \emph{et~al.}
\newblock \bibinfo{title}{Time, space and social interactions: exit mechanisms
  for the covid-19 epidemics}.
\newblock \emph{\bibinfo{journal}{Sci Rep}} \textbf{\bibinfo{volume}{10}},
  \bibinfo{pages}{13764} (\bibinfo{year}{2020}).

\bibitem{Scudellari}
\bibinfo{author}{Scudellari, M.}
\newblock \bibinfo{title}{How the pandemic might play out in 2021 and beyond}.
\newblock \emph{\bibinfo{journal}{Nature}} \textbf{\bibinfo{volume}{584}},
  \bibinfo{pages}{22 -- 25} (\bibinfo{year}{2020}).

\bibitem{Wilson:1971bg}
\bibinfo{author}{Wilson, K.~G.}
\newblock \bibinfo{title}{{Renormalization group and critical phenomena. 1.
  Renormalization group and the Kadanoff scaling picture}}.
\newblock \emph{\bibinfo{journal}{Phys. Rev. B}} \textbf{\bibinfo{volume}{4}},
  \bibinfo{pages}{3174--3183} (\bibinfo{year}{1971}).

\bibitem{Wilson:1971dh}
\bibinfo{author}{Wilson, K.~G.}
\newblock \bibinfo{title}{{Renormalization group and critical phenomena. 2.
  Phase space cell analysis of critical behavior}}.
\newblock \emph{\bibinfo{journal}{Phys. Rev. B}} \textbf{\bibinfo{volume}{4}},
  \bibinfo{pages}{3184--3205} (\bibinfo{year}{1971}).

\bibitem{cacciapaglia2020second}
\bibinfo{author}{Cacciapaglia, G.}, \bibinfo{author}{Cot, C.} \&
  \bibinfo{author}{Sannino, F.}
\newblock \bibinfo{title}{Second wave covid-19 pandemics in europe: A temporal
  playbook}.
\newblock \emph{\bibinfo{journal}{accepted for publication on Sci. Rep}}
  (\bibinfo{year}{2020}).
\newblock \eprint{2007.13100}.

\bibitem{DellaMorte:2020qry}
\bibinfo{author}{Della~Morte, M.} \& \bibinfo{author}{Sannino, F.}
\newblock \bibinfo{title}{{Renormalisation Group approach to pandemics as a
  time-dependent SIR model}}  (\bibinfo{year}{2020}).
\newblock \eprint{2007.11296}.

\bibitem{Kosterlitz:1974sm}
\bibinfo{author}{Kosterlitz, J.}
\newblock \bibinfo{title}{{The Critical properties of the two-dimensional x y
  model}}.
\newblock \emph{\bibinfo{journal}{J. Phys. C}} \textbf{\bibinfo{volume}{7}},
  \bibinfo{pages}{1046--1060} (\bibinfo{year}{1974}).

\bibitem{Miransky:1984ef}
\bibinfo{author}{Miransky, V.}
\newblock \bibinfo{title}{{Dynamics of Spontaneous Chiral Symmetry Breaking and
  Continuum Limit in Quantum Electrodynamics}}.
\newblock \emph{\bibinfo{journal}{Nuovo Cim. A}} \textbf{\bibinfo{volume}{90}},
  \bibinfo{pages}{149--170} (\bibinfo{year}{1985}).

\bibitem{Miransky:1996pd}
\bibinfo{author}{Miransky, V.} \& \bibinfo{author}{Yamawaki, K.}
\newblock \bibinfo{title}{{Conformal phase transition in gauge theories}}.
\newblock \emph{\bibinfo{journal}{Phys. Rev. D}} \textbf{\bibinfo{volume}{55}},
  \bibinfo{pages}{5051--5066} (\bibinfo{year}{1997}).
\newblock \bibinfo{note}{[Erratum: Phys.Rev.D 56, 3768 (1997)]},
  \eprint{hep-th/9611142}.

\bibitem{Holdom:1988gs}
\bibinfo{author}{Holdom, B.}
\newblock \bibinfo{title}{{Raising Condensates Beyond the Ladder}}.
\newblock \emph{\bibinfo{journal}{Phys. Lett. B}}
  \textbf{\bibinfo{volume}{213}}, \bibinfo{pages}{365--369}
  (\bibinfo{year}{1988}).

\bibitem{Holdom:1988gr}
\bibinfo{author}{Holdom, B.}
\newblock \bibinfo{title}{{Continuum Limit of Quenched Theories}}.
\newblock \emph{\bibinfo{journal}{Phys. Rev. Lett.}}
  \textbf{\bibinfo{volume}{62}}, \bibinfo{pages}{997} (\bibinfo{year}{1989}).

\bibitem{Cohen:1988sq}
\bibinfo{author}{Cohen, A.~G.} \& \bibinfo{author}{Georgi, H.}
\newblock \bibinfo{title}{{Walking Beyond the Rainbow}}.
\newblock \emph{\bibinfo{journal}{Nucl. Phys. B}}
  \textbf{\bibinfo{volume}{314}}, \bibinfo{pages}{7--24}
  (\bibinfo{year}{1989}).

\bibitem{Appelquist:1996dq}
\bibinfo{author}{Appelquist, T.}, \bibinfo{author}{Terning, J.} \&
  \bibinfo{author}{Wijewardhana, L.}
\newblock \bibinfo{title}{{The Zero temperature chiral phase transition in
  SU(N) gauge theories}}.
\newblock \emph{\bibinfo{journal}{Phys. Rev. Lett.}}
  \textbf{\bibinfo{volume}{77}}, \bibinfo{pages}{1214--1217}
  (\bibinfo{year}{1996}).
\newblock \eprint{hep-ph/9602385}.

\bibitem{Gies:2005as}
\bibinfo{author}{Gies, H.} \& \bibinfo{author}{Jaeckel, J.}
\newblock \bibinfo{title}{{Chiral phase structure of QCD with many flavors}}.
\newblock \emph{\bibinfo{journal}{Eur. Phys. J. C}}
  \textbf{\bibinfo{volume}{46}}, \bibinfo{pages}{433--438}
  (\bibinfo{year}{2006}).
\newblock \eprint{hep-ph/0507171}.

\bibitem{Sannino:2012wy}
\bibinfo{author}{Sannino, F.}
\newblock \bibinfo{title}{{Jumping Dynamics}}.
\newblock \emph{\bibinfo{journal}{Mod. Phys. Lett. A}}
  \textbf{\bibinfo{volume}{28}}, \bibinfo{pages}{1350127}
  (\bibinfo{year}{2013}).
\newblock \eprint{1205.4246}.

\bibitem{BKT1}
\bibinfo{author}{Hadzibabic, Z.}, \bibinfo{author}{Kr{\"u}ger, P.},
  \bibinfo{author}{Cheneau, M.}, \bibinfo{author}{Battelier, B.} \&
  \bibinfo{author}{Dalibard, J.}
\newblock \bibinfo{title}{Berezinskii--kosterlitz--thouless crossover in a
  trapped atomic gas}.
\newblock \emph{\bibinfo{journal}{Nature}} \textbf{\bibinfo{volume}{441}},
  \bibinfo{pages}{1118 -- 1121} (\bibinfo{year}{2006}).

\bibitem{BKT2}
\bibinfo{author}{Tutsch, U.} \emph{et~al.}
\newblock \bibinfo{title}{Evidence of a field-induced
  berezinskii--kosterlitz--thouless scenario in a two-dimensional spin-dimer
  system}.
\newblock \emph{\bibinfo{journal}{Nature Communications}}
  \textbf{\bibinfo{volume}{5}}, \bibinfo{pages}{5169} (\bibinfo{year}{2014}).

\bibitem{BKT3}
\bibinfo{author}{Situ, G.} \& \bibinfo{author}{Fleischer, J.~W.}
\newblock \bibinfo{title}{Dynamics of the berezinskii--kosterlitz--thouless
  transition in photon fluid}.
\newblock \emph{\bibinfo{journal}{Nature Photonics}}
  \textbf{\bibinfo{volume}{14}}, \bibinfo{pages}{517 -- 522}
  (\bibinfo{year}{2020}).

\bibitem{BKT4}
\bibinfo{author}{Li, H.} \emph{et~al.}
\newblock \bibinfo{title}{Kosterlitz--thouless melting of a magnetic order in
  the triangular quantum ising material tmmggao$_4$}.
\newblock \emph{\bibinfo{journal}{Nature Communications}}
  \textbf{\bibinfo{volume}{11}}, \bibinfo{pages}{1111} (\bibinfo{year}{2020}).

\bibitem{Sannino:2004qp}
\bibinfo{author}{Sannino, F.} \& \bibinfo{author}{Tuominen, K.}
\newblock \bibinfo{title}{{Orientifold theory dynamics and symmetry breaking}}.
\newblock \emph{\bibinfo{journal}{Phys. Rev. D}} \textbf{\bibinfo{volume}{71}},
  \bibinfo{pages}{051901} (\bibinfo{year}{2005}).
\newblock \eprint{hep-ph/0405209}.

\bibitem{Dietrich:2006cm}
\bibinfo{author}{Dietrich, D.~D.} \& \bibinfo{author}{Sannino, F.}
\newblock \bibinfo{title}{{Conformal window of SU(N) gauge theories with
  fermions in higher dimensional representations}}.
\newblock \emph{\bibinfo{journal}{Phys. Rev. D}} \textbf{\bibinfo{volume}{75}},
  \bibinfo{pages}{085018} (\bibinfo{year}{2007}).
\newblock \eprint{hep-ph/0611341}.

\bibitem{Cacciapaglia:2020kgq}
\bibinfo{author}{Cacciapaglia, G.}, \bibinfo{author}{Pica, C.} \&
  \bibinfo{author}{Sannino, F.}
\newblock \bibinfo{title}{{Fundamental Composite Dynamics: A Review}}.
\newblock \emph{\bibinfo{journal}{in press on Phys. Reports}}
  (\bibinfo{year}{2020}).
\newblock \eprint{2002.04914}.

\end{thebibliography}

\begin{itemize}
 \item[] {\bf Acknowledgements:} G.C.  acknowledges partial support from the Labex-LIO (Lyon Institute of Origins) under grant ANR-10-LABX-66 (Agence Nationale pour la Recherche), and FRAMA (FR3127, F\'ed\'eration de Recherche ``Andr\'e Marie Amp\`ere''). 
 \item[] {\bf Author Contribution:} This work has been designed and performed conjointly and equally by the authors. G.C. and F.S. have equally contributed to the writing of the article.
 \item[] {\bf Competing Interests:} The authors declare that they have no competing financial and non-finantial interests.
 \item[] {\bf Correspondence:} Correspondence and requests for materials should be addressed to G.Cacciapaglia~(email: g.cacciapaglia@ipnl.in2p3.fr).
\end{itemize}

\end{document}